\documentstyle[12pt]{article}

\input{tcilatex}
\begin{document}

\title{{\Large Superconformal Field Theory with Boundary:}\\
Spin Model}
\author{S.A.Apikyan $^{a,b}$ $^\dagger$, D.A.Sahakyan $^b$ $\ddagger$ \\
{\small $^{a)}$ Yerevan Physics Institute}\\
{\small Alikhanian 2, Yerevan, 375036 Armenia.}\\
{\small $^{b)}$ Yerevan State University,}\\
{\small Al.Manoogian 1, Yerevan, 375049 Armenia.}}
\maketitle

\begin{abstract}
GSO projected Superconformal field theory (Spin Model) with boundary is
considered. There were written the boundary states. For this model were
derived one-point structure constants and "bootstrap" equations for
boundary-bulk structure constants.
\end{abstract}

\vfill

\hrule

\noindent
$^{\dagger }$ e-mail: apikyan@lx2.yerphi.am\newline
$^{\ddagger }$ e-mail: sahakian@uniphi.yerphi.am

\section{\bf Introduction}

Superconformal Field Theory on manifold with boundary plays an important
role in open superstring theories and are the basic ingredient for the
construction of the open superstring theory. Perhaps it can be also
essential for some two dimensional exactly solvable models and their
critical phenomenon.

Here we recall basic facts from the superconformal field theory adapted to
our case and establish our notation. The basic ideas of superconformal field
theory can be found in refs \cite{FQSh},\cite{BKT}.

Supersymmetric extensions of Virasoro algebra are obtained by generalizing
conformal transformations to superconformal transformations of
supercoordinates $\hat{z}=(z,\theta )$. The generators of superconformal
transformations 
\begin{equation}
\begin{array}{ll}
\delta z=u+\theta \epsilon ; & \delta \theta =\epsilon +\frac 12\theta u_z;
\\ 
\delta \bar{z}=\bar{u}+\bar{\theta}\bar{\epsilon}; & \delta \bar{\theta}=%
\bar{\epsilon}+\frac 12\bar{\theta}\bar{u}_{\bar{z}}
\end{array}
\label{eq:ST}
\end{equation}
are super stress-energy tensor $G(z,\theta )={\frac 12}S(z)+\theta T(z)$.
The operators $L_n$ and $S_r$ (Laurent coefficients of $T$ and $S$) generate
analytic coordinate and supersymmetry transformations respectively and obey
the algebra, 
\begin{equation}
\begin{array}{l}
\lbrack L_m,L_n]=(m-n)L_{m+n}+\frac c8m(m^2-1)\delta _{m+n} \\ 
\{S_r,S_s\}=2L_{r+s}+\frac c2(r^2-\frac 14)\delta _{r+s} \\ 
\lbrack L_m,S_r]=(\frac m2-r)S_{m+r}
\end{array}
\label{eq:AL}
\end{equation}
The algebra has a $Z_2$ symmetry, so there are two possible moding for the
fermionic generator $S_r$, either half-integer $(r\in 1/2+Z)$ or integer $%
(r\in Z)$ giving the Neveu-Schwarz (${\bf NS}$) and Ramond (${\bf R}$)
algebras respectively. Highest weight states $\mid h\rangle $ of the $NS$
and $R$ algebras satisfy 
\begin{equation}
\begin{array}{l}
L_n\mid h\rangle =S_r\mid h\rangle =0,\qquad n,r>0 \\ 
L_0\mid h\rangle =h\mid h\rangle
\end{array}
\label{eq:Arm}
\end{equation}
Representation are built up by applying the raising operators $L_n$,$S_r$
with $n,r>0$ to the highest weight state $\mid h\rangle $. In the Ramond
sector superconformal current has zero mode, which form two dimensional
Clifford Algebra with the Fermion Number Operator $\Gamma =(-)^F$, commuting
with the $L_0$. As a result, we have double degeneration of the ground state 
\cite{FQSh}. In this space we can choose the following ortogonal basis $\mid
h^{+}\rangle ={R_h}^{+}(0)\mid 0\rangle ,\,\mid h^{-}\rangle ={R_h}%
^{-}(0)\mid 0\rangle $ (where $R^{\pm }$-Ramond spin fields): 
\begin{equation}
\mid h^{-}\rangle =S_0\mid h^{+}\rangle  \label{eq:hS}
\end{equation}
where $\mid h^{+}\rangle $ and $\mid h^{-}\rangle $ are eigenvectors of
operator $(-)^F$ with eigenvalues $+1$ and $-1$ respectively having the same
conformal weight $h$. Using commutation relations (\ref{eq:AL}) we can
obtain: 
\begin{equation}
S_0\mid h^{-}\rangle =S_0^2\mid h^{+}\rangle =(L_0-\frac c{16})\mid
h^{+}\rangle =(h-\frac c{16})\mid h^{+}\rangle  \label{eq:SL}
\end{equation}
Thus, if one normalizes $\mid h^{+}\rangle $ as, $\langle h^{+}\mid
h^{+}\rangle =1$ , then from (\ref{eq:SL}) it follows, that $\langle
h^{-}\mid h^{-}\rangle =h-\frac c{16}$. In case, when $h\ne \frac c{16}$, it
can be chosen basis $\mid {h^{\prime }}^{\pm }\rangle $ such, that $S_0\mid {%
h^{\prime }}^{\pm }\rangle =\sqrt{h-\frac c{16}}\mid {h^{\prime }}^{\mp
}\rangle $ and which is ortonormal. In further we will use the basis (\ref
{eq:hS}). Let us note, that if $h=\frac c{16}$, then $\mid h^{-}\rangle $
becomes 0-vector and decouples from representation of algebra. Hence chiral
symmetry of the ground state is destroyed and the global supersymmetry is
restored.

In the general superconformal theory the full operator algebra of $NS$
superfields and $R^{\pm}$ spin fields is nonlocal \cite{FQSh}. There are two
possibility for projecting onto a local set of fields. First one, keeping
only the $NS$-sector giving the usual algebra of superfields, a fermionic
model. The second one, we can get a local field theory the "spin model"
restricting in superconformal field theory by $\Gamma=1$ sector. In this
paper we are going to consider "Spin Model" with boundary defined on the
upper half plane (the "Fermionic Model" there was studied in ref.\cite{AS}).

It is easy to see, that the requirement of preservation of the geometry
gives strong limitations on parameters of superconformal transformation. One
can see that the expansion coefficients of parameters must be real.
Therefore holomorphic and anti--holomorphic transformations are not
independent. So, let's make analytical continuation of $T$ and $S$ on to
lower half plane. 
\begin{equation}
T(z)=\bar{T}(z);\qquad S(z)=\bar{S}(z);\qquad for\quad Imz<0  \label{eq:AN}
\end{equation}
It means that now we have only one algebra (\ref{eq:AL}) in opposite to
''bulk'' theory, there were two, holomorphic and anti--holomorphic algebras,
which is consistent with the fact, that in theory with boundary we have only
one set of coefficient in expansion of parameters.. Then for $\langle
X\rangle =\langle R^{\pm }(z_1,\bar{z}_1)...R^{\pm }(z_n,\bar{z}_n)\rangle $
correlation function from (\ref{eq:AN}) (using bulk OPE) follows, that in
contrast to bulk Ward Identity where $T(z)$ and $S(z)$ acts only on $%
(z_1,...,z_n)$, in theory with boundary the action of $T(z)$ and $S(z)$ is
extended to $(z_1,\bar{z}_1,...,z_n,\bar{z}_n)$ and hence, in the relations
of the boundary Ward Identity the doubling of terms on the right hand sides
takes place due to terms with $z_i^{\prime }=\bar{z}_i$. So, correlation
function for Ramond fields $\langle X(z_1,\bar{z}_1,...z_n,\bar{z}_n)\rangle
_B$ in our geometry satisfies the same differential equation as does bulk
correlation function of Ramond fields $\langle X(z_1,\bar{z}_1,...z_{2n},%
\bar{z}_{2n})\rangle $.

\vspace{0.4cm}

\section{Boundary States}

Further we will construct boundary states of theories defined on the upper
half plane or strip, which one can also interpretate as a world sheet of an
open superstring. Mapping of the upper half plane on to strip is given by
the conformal transformation $z=e^{t+i\sigma }$, where $(t,\sigma )$ are
coordinates on strip $(0,\pi )$.

In general superconformal field theory with boundary, the unique requirement
on boundary condition is the superconformal invariance: 
\begin{equation}
\begin{array}{l}
T(z=e^t)=\bar{T}(\bar{z}=e^t)\qquad T(z=e^{t+i\pi })=\bar{T}(\bar{z}%
=e^{t-i\pi }) \\ 
S(z=e^t)=\bar{S}(\bar{z}=e^t)\qquad S(z=e^{t+i\pi })=\bar{S}(\bar{z}%
=e^{t-i\pi })\quad NS-sector \\ 
S(z=e^t)=\bar{S}(\bar{z}=e^t)\qquad S(z=e^{t+i\pi })=-\bar{S}(\bar{z}%
=e^{t-i\pi })\quad R-sector
\end{array}
\label{eq:TS}
\end{equation}
If one compactifies $t$ by mod $2\pi Im\tau $ ($\tau $ is purely imaginary)
he obtaines the theory defined on a cylinder with radius $Im\tau $. Then
partition functions with boundary conditions $\alpha ,\beta $ at the ends of
cylinder can be written (for antiperiodic and periodic boundary condition in
time direction) as follows, 
\begin{equation}
\begin{array}{ll}
\vspace{0.3cm}Z_{\alpha \beta }^{NS}=Tre^{2\pi i\tau H_{\alpha \beta
}^{open}};\quad & Z_{\alpha \beta }^{(-)NS}=Tr(-1)^Fe^{2\pi i\tau H_{\alpha
\beta }^{open}} \\ 
\vspace{0.3cm}Z_{\alpha ^{\prime }\beta ^{\prime }}^R=Tre^{2\pi i\tau
H_{\alpha ^{\prime }\beta ^{\prime }}^{open}}; & Z_{\alpha ^{\prime }\beta
^{\prime }}^{(-)R}=Tr(-1)^Fe^{2\pi i\tau H_{\alpha ^{\prime }\beta ^{\prime
}}^{open}}
\end{array}
\label{eq:1}
\end{equation}
The bulk superconformal algebra is the tensor product of two algebras,
therefore natural chirality operator is $\Gamma =(-1)^{F_{tot}}$, where $%
F_{tot}=F+\bar{F}$ is the fermion number of the full algebra. The projection
of boundary SCFT is analogous to the GSO projection of the bulk SCFT with
the difference that in the boundary theory only one chirality operator $%
\Gamma =(-1)^F$ is defined, since for boundary case there is just one
algebra. The projection to local theory in $NS$ and $R$ sectors is given by $%
\Gamma =1$.

Let's note that summarizing partition functions in each sector $%
Z_{\alpha\beta}^{NS}+Z_{\alpha\beta}^{(-)NS}$ and $Z_{\alpha^{\prime}\beta^{%
\prime}}^R+Z_{\alpha^{\prime}\beta^{\prime}}^{(-)R}$ are just projecting
into subspace having even fermion number.

From the other side, the same partition function can be considered as a
propagation of closed superstring on $\sigma $ direction between boudary
states $\langle \alpha \mid ,\mid \beta \rangle $, 
\begin{equation}
Z_{\alpha \beta }=\langle \alpha \mid e^{-\pi H^{cyl}}\mid \beta \rangle
=\langle \alpha \mid e^{-\frac \pi {Im\tau }(L_0^{cyl}+\bar{L}_0^{cyl})}\mid
\beta \rangle  \label{eq:2}
\end{equation}
where $H^{cyl}$ is the Hamiltonian for closed superstring, $L_0^{cyl}$, ${%
\bar{L}}_0^{cyl}$ are generators of Virasoro and $\mid \alpha \rangle $, $%
\mid \beta \rangle $ satisfy to conditions (\ref{eq:TS}), which can be
rewritten as 
\begin{equation}
\begin{array}{lll}
\vspace{0.3cm}T^{cyl}(\zeta )={\bar{T}}^{cyl}(\bar{\zeta})|_{\zeta =e^{-it}}
& T^{cyl}(\zeta )=\bar{T}^{cyl}(\bar{\zeta})|_{\zeta =e^{\pi -it}} &  \\ 
\vspace{0.3cm}S^{cyl}(\zeta )=-i\bar{S}^{cyl}(\bar{\zeta}|_{\zeta =e^{-it}}
& S^{cyl}(\zeta )=-i\bar{S}^{cyl}(\bar{\zeta})|_{\zeta =e^{\pi -it}} & (R)
\\ 
\vspace{0.3cm}S^{cyl}(\zeta )=-i\bar{S}^{cyl}(\bar{\zeta})|_{\zeta =e^{-it}}
& S^{cyl}(\zeta )=i\bar{S}^{cyl}(\bar{\zeta})|_{\zeta =e^{\pi -it}} & (NS)
\end{array}
\label{eq:zeta}
\end{equation}
where $\zeta =e^{-i(t+i\sigma )}$. One can rewrite conditions (\ref{eq:zeta}%
), in the form: 
\begin{equation}
\begin{array}{l}
(L_n-\bar{L}_{-n})\mid B_{\pm }\rangle =0 \\ 
(S_r\pm i\bar{S}_{-r})\mid B_{\pm }\rangle =0
\end{array}
\label{eq:B}
\end{equation}
where $r\in Z$ or $r\in Z+\frac 12$.

\smallskip It is easy to see from (\ref{eq:zeta}),(\ref{eq:B}) that one
should choose ``$+$'' boudary states (or ``$-$'') for both ends of the
cylinder for propagation of Neveu-Schwarz and ``$+-$'' (or ``$-+$'' ) for
propagation of Ramond states in open string channel.. Of course ``$+$'' and
``$-$'' states are not essentially different. For our purposes we will fix ``%
$+$'' boundary states for $\sigma =0$ end of cylinder and vary ``$+$'' and ``%
$-$'' for the other end.

One of the basic aims of this paper is to find solutions (\ref{eq:B}) in
each irreducable representation of superconformal algebra. The solution of
conditions (\ref{eq:B}) in $NS$ sector is given by the following anzats \cite
{Is}, 
\begin{equation}
\mid h_{\pm }^{NS}\rangle =\sum_{s\in Z^{+}/2}\mid h,s\rangle \otimes U_{\pm
}^{NS}\overline{\mid h,s\rangle }  \label{eq:sum}
\end{equation}
where $U_{NS}^{\pm }$ is an anti-unitary operators, satisfying the following
conditions: 
\begin{equation}
\begin{array}{l}
L_nU_{\pm }^{NS}=U_{\pm }^{NS}L_n \\ 
U_{\pm }^{NS}S_r=\mp iS_rU_{\pm }^{NS}(-)^F
\end{array}
\label{eq:AP}
\end{equation}
One can see that equations (\ref{eq:AP}) yield

\begin{equation}
U_{\pm }^{NS}\mid h,s\rangle =\frac{1-i}2(1\pm i(-)^F)\mid h,s\rangle
\label{AP'}
\end{equation}

It's easy to show, that (\ref{eq:sum}) satisfies to conditions (\ref{eq:B}).
For this purpose we just have to check, that for any basic vector $\overline{%
\langle i\mid }\otimes \langle j\mid $, following relations are valid, 
\begin{equation}
\begin{array}{l}
\overline{\langle i\mid }\otimes \langle j\mid (L_n-\bar{L}_{-n})\mid h_{\pm
}^{NS}\rangle =0, \\ 
\overline{\langle i\mid }\otimes \langle j\mid (S_r\pm i\bar{S}_{-r})\mid
h_{\pm }^{NS}\rangle =0.
\end{array}
\end{equation}
It is more interesting Ramond sector. For the beginning let us consider the
case $h\ne c/16$. We can use the same anzats (\ref{eq:sum}) to solve (\ref
{eq:B}), 
\begin{eqnarray}
\vspace{0.3cm} &\mid &h_{\pm }^R\rangle =\sum_{q\in Z^{+}}\mid h,q\rangle
\otimes U_{\pm }^R\overline{\mid h,q\rangle }=  \nonumber \\
\sum_{p\in N} &\mid &h^{+},p\rangle \otimes U_{\pm }^R\overline{\mid
h^{+},p\rangle }+\sum_{p\in N}\mid h^{-},p\rangle \otimes U_{\pm }^R%
\overline{\mid h^{-},p\rangle }
\end{eqnarray}
where $U_{\pm }^R$ is anti-unitary operators, satisfying to conditions: 
\begin{equation}
\begin{array}{l}
L_nU_{\pm }^R=U_{\pm }^RL_n \\ 
U_{\pm }^RS_r=\mp iS_rU_{\pm }^R(-)^F
\end{array}
\end{equation}
Since the ground state is now non--trivial, we have freedom in a definition
of the action $U_{\pm }^R$ on this space. And we have the only restriction
on $U_{\pm }^R$: 
\begin{equation}
(U_{\pm }^RS_0\pm iS_0U_{\pm }^R(-)^F)\mid h^{\pm }\rangle =0  \label{eq:yr}
\end{equation}
In representation, where 
\[
\mid h^{+}\rangle ={\binom 10}\quad and\quad \mid h^{-}\rangle ={\binom 0{%
\sqrt{h-\frac c{16}}}} 
\]
$S_0$ and $(-)^F$ can be represented as 
\begin{equation}
\begin{array}{l}
S_0=\sqrt{h-\frac c{16}}\sigma _x;\qquad (-)^F=\sigma _z
\end{array}
\label{eq:ug}
\end{equation}
where $\sigma _x$ and $\sigma _z$ are Pauli matrixes. Using (\ref{eq:yr})
and representation (\ref{eq:ug}), we get: 
\begin{equation}
U_{\pm }^R=\left( 
\begin{array}{cc}
a & \mp ic \\ 
c & \mp ia
\end{array}
\right)
\end{equation}
where $a$ and $c$ satisfy anti-unitary condition: $aa^{*}+cc^{*}=1$ and $%
ac^{*}+a^{*}c=0$. According to latter equations there are two independent
choices for $U_{\pm }^R$:

\begin{equation}
\begin{array}{lll}
U_{\pm }^R=\left( 
\begin{array}{cc}
1 & 0 \\ 
0 & \mp i
\end{array}
\right) & \text{or} & U_{\pm }^R=\left( 
\begin{array}{cc}
0 & \mp i \\ 
1 & 0
\end{array}
\right)
\end{array}
\end{equation}
It is interesting to note, that for $h=\frac c{16}$, the uniqueness of $%
U_{\pm }^R$ is recovered. The nature of this degeneration is very
interesting but we will not analyze it. We only note that it is sufficient
to restrict to first choice of $U_{\pm }^R$.

The partition functions (\ref{eq:1}) of the theory defined on compactified
cylinder can be expressed as a linear combination of characters since
instead of holomorphic and atiholomorphic algebras (in the bulk) now there
is just one algebra: 
\begin{equation}
\begin{array}{ll}
\vspace{0.3cm}Z_{\alpha \beta }^{NS}=\sum n_{\alpha \beta }^i\chi
_i^{NS}(q),\qquad  & Z_{\alpha \beta }^{(-)NS}=\sum n_{\alpha \beta }^i\chi
_i^{(-)NS}(q); \\ 
\vspace{0.3cm}Z_{\alpha \beta }^R=\sum m_{\alpha \beta }^i\chi _i^R(q), & 
Z_{\alpha \beta }^{(-)R}=\sum m_{\alpha \beta }^i\chi _i^{(-)R}
\end{array}
\end{equation}
here $\chi _i^{NS}(q)=q^{-\hat{c}/16}Tr_iq^{L_0}$, $\chi _i^{(-)NS}(q)=q^{-%
\hat{c}/16}Tr_i(-1)^Fq^{L_0}$ and  $\chi _i^R(q)=q^{-\hat{c}/16}Tr_iq^{L_0}$%
, $\chi _i^{(-)R}=Tr_i(-1)^F$ are the characters of the superconformal
algebras in $NS$ and $R$ sectors respectively. For the last character note
that $R$ fermion has zero energy on the cylinder at the supersymmetric
ground state $(h=\hat{c}/16)$. By non-negative integer $n_{\alpha \beta }^i$%
, $m_{\alpha \beta }^i$ denoted the number of times that representation $i$
occurs in the spectrum of $H_{\alpha \beta }^{open}$.

The character formulas for the $NS$ and $R$ algebra have been derived by
Goddard, Kent and Olive \cite{GKO} and by Kac and Wakimoto \cite{KW} and
under the modular transformation $\tau \rightarrow -1/\tau $ the characters
for the ''spin model'' transform linearly \cite{K}, 
\begin{equation}
\begin{array}{l}
\chi _i^{NS}(q)=\sum (S_{NS}^{NS})_i^j\chi _j^{NS}(\tilde{q}), \\ 
\chi _i^{(-)NS}(q)=\sum (S_R^{NS})_i^j\chi _j^R(\tilde{q}), \\ 
\chi _i^R(q)=\sum (S_{NS}^R)_i^j\chi _j^{(-)NS}(\tilde{q})
\end{array}
\end{equation}
which leads to 
\begin{equation}
\begin{array}{l}
Z_{\alpha \beta }^{NS}=\sum n_{\alpha \beta }^i(S_{NS}^{NS})_i^j\chi _j^{NS}(%
\tilde{q}) \\ 
Z_{\alpha \beta }^{(-)NS}=\sum n_{\alpha \beta }^i(S_R^{NS})_i^j\chi _j^R(%
\tilde{q}) \\ 
Z_{\alpha \beta }^R=\sum m_{\alpha \beta }^i(S_{NS}^R)_i^j\chi _j^{(-)NS}(%
\tilde{q}) \\ 
Z_{\alpha \beta }^{(-)R}=\sum m_{\alpha \beta }^i\chi _i^{(-)R}
\end{array}
\label{eq:3}
\end{equation}
where $\tilde{q}=e^{-2\pi i/\tau }$. In order to have complete set of
boundary states defined by equation (\ref{eq:sum}), we have to consider
diagonal bulk theory. Following to Cappelli and Kastor \cite{K} there are
different superconformal theories corresponding to different modular
invariant combination of characters 
\begin{equation}
Z_{NS,R}=\sum_{i,j}F_{i,j}N_{i,j}\chi _i(q)\bar{\chi}_j(\bar{q})
\end{equation}
here the factor $F$ is equal to 2 for the nonsupersymmetric $R$ highest
weight states, which one twofold degenerated, and is equal to 1 otherwise. $%
N_{i,j}$ is the number of highest weight states $(h_i,\bar{h}_j)$ in the
bulk theory which one obeys to the sum rules, requiring modular invariance
of $Z_{NS}(q)=Z_{NS}(\tilde{q})$, $Z_R(q)=Z_{NS}^{(-)}(\tilde{q})$, $%
Z_{NS}^{(-)}(q)=Z_R(\tilde{q})$ we can get 
\begin{equation}
\begin{array}{l}
\vspace{0.3cm}\sum N_{nm,kl}\sin \frac{\pi nn^{\prime }}p\sin \frac{\pi
mm^{\prime }}{p+2}\sin \frac{\pi kk^{\prime }}p\sin \frac{\pi ll^{\prime }}{%
p+2}=\frac{p(p+2)}{16}N_{n^{\prime }m^{\prime },k^{\prime }l^{\prime }} \\ 
\vspace{0.3cm}\sum N_{nm,kl}Y_{nm,kl}\sin \frac{\pi nn^{\prime }}p\sin \frac{%
\pi mm^{\prime }}{p+2}\sin \frac{\pi kk^{\prime }}p\sin \frac{\pi ll^{\prime
}}{p+2}=\frac{(-1)^\alpha p(p+2)}{16}N_{n^{\prime }m^{\prime },k^{\prime
}l^{\prime }} \\ 
\vspace{0.3cm}\sum N_{nm,kl}(-1)^\alpha \sin \frac{\pi nn^{\prime }}p\sin 
\frac{\pi mm^{\prime }}{p+2}\sin \frac{\pi kk^{\prime }}p\sin \frac{\pi
ll^{\prime }}{p+2}=\frac{p(p+2)}{16}Y_{n^{\prime }m^{\prime },k^{\prime
}l^{\prime }}^{-1}N_{n^{\prime }m^{\prime },k^{\prime }l^{\prime }}
\end{array}
\end{equation}
There are at least two series of solutions to the above sum rules. One of
these the diagonal (or scalar) solution of the superconformal sum rules are
given by $N_{nm,kl}=\delta _{nk}\delta _{ml}$ in $NS,R$ sectors. For the
diagonal theory the constructed states $\mid j_{\pm }^{R,NS}\rangle $ give
complete set in the space of all boundary states and we can therefore write 
\begin{equation}
\mid \alpha _{\pm }\rangle =\sum_j\mid j_{\pm }\rangle \langle j_{\pm }\mid
\alpha _{\pm }\rangle =\sum_j\mid j_{\pm }^{NS}\rangle \langle j_{\pm
}^{NS}\mid \alpha _{\pm }\rangle +\sum_j\mid j_{\pm }^R\rangle \langle
j_{\pm }^R\mid \alpha _{\pm }\rangle 
\end{equation}
Using these representations we can rewrite (\ref{eq:2}) 
\begin{eqnarray}
Z_{\alpha _{+}\beta _{+}}(\tilde{q}) &=&\sum \langle \alpha _{+}\mid
i_{+}^{NS}\rangle \langle i_{+}^{NS}\mid \beta _{+}\rangle \chi _i^{NS}(%
\tilde{q})+\sum \langle \alpha _{+}\mid i_{+}^R\rangle \langle i_{+}^R\mid
\beta _{+}\rangle \chi _i^R(\tilde{q})  \nonumber \\
Z_{\alpha _{-}\beta _{+}}(\tilde{q}) &=&i\sum \langle \alpha _{-}\mid
i_{-}^{NS}\rangle \langle i_{+}^{NS}\mid \beta _{+}\rangle \chi _i^{(-)NS}(%
\tilde{q})+i\sum \langle \alpha _{-}\mid i_{-}^R\rangle \langle i_{-}^R\mid
\beta _{+}\rangle \chi _i^{(-)R}  \nonumber \\
&&
\end{eqnarray}
For such theories, when each representation occurs just once in the spectrum
of bulk $H$, we have linearly independent different characters, therefore
comparing last relations and (\ref{eq:3}), namely $Z_{\alpha _{+}\beta _{+}}(%
\tilde{q})=Z_{\alpha \beta }^{NS}(q)+Z_{\alpha \beta }^{(-)NS}(q)$ and $%
Z_{\alpha _{-}\beta _{+}}(\tilde{q})=Z_{\alpha \beta }^R(q)+Z_{\alpha \beta
}^{(-)R}(q)$ we can get immediently relations

\begin{eqnarray}
\sum_j(S_{NS}^{NS})_j^in_{\alpha _{+}\beta _{+}}^j &=&\langle \alpha
_{+}\mid i_{+}^{NS}\rangle \langle i_{+}^{NS}\mid \beta _{+}\rangle  
\nonumber \\
\sum_j(S_R^{NS})_j^in_{\alpha _{+}\beta _{+}}^j &=&\langle \alpha _{+}\mid
i_{+}^R\rangle \langle i_{+}^R\mid \beta _{+}\rangle   \label{6} \\
\sum_j(S_{NS}^R)_j^im_{\alpha _{+}\beta _{-}}^j &=&i\langle \alpha _{+}\mid
i_{+}^{NS}\rangle \langle i_{-}^{NS}\mid \beta _{-}\rangle   \nonumber \\
m_{\alpha _{+}\beta _{-}}^{h=c/16} &=&i\langle \alpha _{+}\mid h=c/16\rangle
\langle h=c/16\mid \beta _{-}\rangle   \nonumber
\end{eqnarray}

Thus, solving equations for coefficients of boundary states $|\alpha
_{+}\rangle $ (in the same way for $|\alpha _{-}\rangle $), we can write
finaly particulary for $|\alpha _{+}\rangle $ following expression 
\begin{equation}
\mid \tilde{k}\rangle =\sum_j\frac{(S_{NS}^{NS})_k^j}{%
[(S_{NS}^{NS})_0^j]^{1/2}}\mid j_{+}^{NS}\rangle +\sum_j\frac{(S_R^{NS})_k^j%
}{[(S_R^{NS})_0^j]^{1/2}}\mid j_{+}^R\rangle  \label{eq:5}
\end{equation}
These states have property that $n_{\tilde{0}\tilde{k}}^i=\delta _k^i$ which
means that the representation $k$ appears in the spectrum of $H_{\tilde{0}%
\tilde{k}}$.

\section{One and three point boundary correlation functions}

In the superstring theories we generally are interested in calculation of
scattering amplitudes with both open and closed strings in the initial and
final states. A string diagram with external open and closed string can be
conformally mapped to the upper half plane. After this mapping the external
open string are represented by vertex operators at finite points on the
boundary, while the closed strings are represented by vertex operators at
finite points on the upper half plane. All of this means that for
construction open and closed superstring theories we are really interested
in superconformal field theory with boundary (SCFT on half plane). One of
the interesting question is how in the intermediate channel of string
diagram (with external open and closed strings) closed string vertex can be
expressed by open string vertex operators with given type of boundary
condition. In a superconformal field theory (in which the boundary
conditions do not break the superconformal symmetry) this can be represented
as short distance expansion of bulk vertex operators near a boundary \cite
{AS}. There are two types of bulk fields: Ramond spin fields $R(z,\bar{z})$
and Neveu-Scwarz superfields 
\begin{equation}
\Phi (\hat{z},\bar{z})=\phi (z,\bar{z})+\theta \Psi (z,\bar{z})+\bar{\theta}%
\bar{\Psi}(z,\bar{z})+\theta \bar{\theta}F(z,\bar{z})
\end{equation}
where 
\begin{equation}
\begin{array}{l}
\Psi (z,\bar{z})=S_{-1/2}\phi (z,\bar{z}); \\ 
\bar{\Psi}(z,\bar{z})=\bar{S}_{-1/2}\phi (z,\bar{z}); \\ 
F(z,\bar{z})=S_{-1/2}\bar{S}_{-1/2}\phi (z,\bar{z})
\end{array}
\label{set}
\end{equation}
one can write short distance expansion for $\phi (z,\bar{z})$ and $R(z,\bar{z%
})$ near boundary as follows 
\begin{eqnarray}
\phi (z,\bar{z}) &=&\sum_i(z-\bar{z})^{\Delta _{\phi _i}^B-\Delta _\phi
}C_{\phi \phi _i^B}^B[\phi _i^B(x)] \\
R(z,\bar{z}) &=&\sum_i(z-\bar{z})^{\Delta _{\phi _i}^B-\Delta _R}C_{R\phi
_i^B}^B[\phi _i^B(x)]  \label{OPE}
\end{eqnarray}
here $[\phi ^B(x)]$, --are conformal class of boundary vertex operators $%
\phi ^B$ and $C_{\phi \phi ^B}^B$, $C_{R\phi ^B}^B$--are OPE's boundary
structure constants of Neveu-Schwarz and Ramond fields respectively. From (%
\ref{set}) it is possible to obtain corresponding relations for $\Psi $ and $%
F$ fields.

Now let's obtain these boundary structure constants. First of all note that
for identity boundary operator corresponding structure constant is equal to
constant factor of one point boundary correlation function. One point
boundary correlation (with boundary conditions labelled by B) of $NS$ and $R$
fields with corresponding to superconformal invariance and boundary Ward
identity can be written 
\begin{eqnarray}
\langle \Phi (\hat{z},\bar{z})\rangle _B &=&\frac{A_\Phi ^B}{(z-\bar{z}%
-\theta \bar{\theta})^{\Delta _\phi }}  \nonumber \\
\left\langle R^{+}(z,\bar{z})\right\rangle &=&\frac{A_R^B}{(z-\bar{z}%
)^{\Delta _R}}  \label{one}
\end{eqnarray}
where $A_\Phi ^B=C_{\Phi I}^B$, $A_R^B=C_{RI}^B$ . It is easy to see that $%
A_\Phi ^B=A_\phi ^B$. Thus, according to the definition \cite{C} \cite{CL}, 
\begin{equation}
A_\phi ^B=\frac{\langle \phi \mid B\rangle }{\langle 0\mid B\rangle }\qquad
A_R^B=\frac{\langle R^{+}\mid B\rangle }{\langle 0\mid B\rangle }
\end{equation}
and using the superconformal physical boundary states (\ref{eq:5}) we find 
\begin{eqnarray}
A_\phi ^{\tilde{k}} &=&\frac{\langle \phi \mid \tilde{k}\rangle }{\langle
0\mid \tilde{k}\rangle }=\frac{[(S_{NS}^{NS})_0^0]^{1/2}}{%
[(S_{NS}^{NS})_0^\phi ]^{1/2}}\frac{(S_{NS}^{NS})_k^\phi }{(S_{NS}^{NS})_k^0}
\nonumber \\
A_R^{\tilde{k}} &=&\frac{\langle R^{+}\mid \tilde{k}\rangle }{\langle 0\mid 
\tilde{k}\rangle }=\frac{[(S_{NS}^{NS})_0^0]^{1/2}}{[(S_R^{NS})_0^R]^{1/2}}%
\frac{(S_R^{NS})_k^R}{(S_{NS}^{NS})_k^0}
\end{eqnarray}
To determine the boundary structure constants $C_{\phi \phi ^B}^B$, $%
C_{R\phi ^B}^B$ we use associativity of the boundary operator algebra which
imposes global constraints on correlation functions. For this purpose,
consider 2-point functions, 
\begin{equation}
\langle \phi _i(z_1,\bar{z}_1)\phi _j(z_2,\bar{z}_2)\rangle _B;\qquad
\langle R_i(z_1,\bar{z}_1)R_j(z_2,\bar{z}_2)\rangle _B  \label{two}
\end{equation}
in two channels. Of course corresponding correlation functions for $\Psi $
and $F$ can be restored from (\ref{two}) by supersymmetry. We can evaluate
these correlation functions using OPE in different crossing channels.
Associativity of the operator algebra implies that correlation function of
these two channels should give the same result (crossing symmetry), 
\begin{equation}
\sum_kC_{\phi _i\phi _k^B}^{\tilde{n}}C_{\phi _j\phi _k^B}^{\tilde{n}%
}F_{ii;jj}^k(1-\eta )=\sum_mC_{ijm}A_{\phi _m}^{\tilde{n}}F_{ij;ij}^m(\eta )
\label{eq:6}
\end{equation}
\begin{equation}
\sum_kC_{R_\rho \phi _k^B}^{\tilde{n}}C_{R_\sigma \phi _k^B}^{\tilde{n}%
}F_{\rho \rho ;\sigma \sigma }^k(1-\eta )=\sum_mC_{\rho \sigma m}A_{\phi
_m}^{\tilde{n}}F_{\rho \sigma ;\rho \sigma }^m(\eta )  \label{eq:7}
\end{equation}
here $\eta =\mid z_1-z_2\mid ^2/\mid z_1-\bar{z}_2\mid ^2$ is cross-ratios, $%
F_{ij;ij}^m(\eta )$, $F_{\rho \sigma ;\rho \sigma }^m(\eta )$, $C_{ijm}$ and 
$C_{\rho \sigma m}$ are conformal blocks and bulk structure constants
respectiviely. According to different basis of differential equations (to
which obey conformal blocks) the solutions are expressed by each other
linearly \cite{Kit}, 
\begin{eqnarray}
F_{ij;ij}^k(\eta ) &=&\sum \alpha \left[ 
\begin{array}{ll}
i & i \\ 
j & j
\end{array}
\right] _{kl}^{NS}F_{ii;jj}^l(1-\eta )  \nonumber \\
F_{\rho \sigma ;\rho \sigma }^k(\eta ) &=&\sum \alpha \left[ 
\begin{array}{ll}
\rho & \rho \\ 
\sigma & \sigma
\end{array}
\right] _{kl}^RF_{\rho \rho ;\sigma \sigma }^l(1-\eta )  \label{boot}
\end{eqnarray}
using above equations from (\ref{eq:6}-\ref{eq:7}) we obtain immediately 
\begin{equation}
C_{\phi _i\phi _l^B}^{\tilde{k}}C_{\phi _j\phi _l^B}^{\tilde{k}%
}=\sum_mC_{ijm}A_{\phi _m}^{\tilde{k}}\alpha \left[ 
\begin{array}{ll}
i & i \\ 
j & j
\end{array}
\right] _{ml}^{NS}
\end{equation}
\begin{equation}
C_{R_\rho \phi _l^B}^{\tilde{k}}C_{R_\sigma \phi _l^B}^{\tilde{k}%
}=\sum_mC_{\rho \sigma m}A_{\phi _m}^{\tilde{k}}\alpha \left[ 
\begin{array}{ll}
\rho & \rho \\ 
\sigma & \sigma
\end{array}
\right] _{ml}^R
\end{equation}
So, all boundary structure constants are expressed via well known bulk
quantities.

\vspace{0.5cm}

This work was partially supported by the INTAS foundation under grant
96-0482. We thank the ICTP (Trieste) for its hospitality where this work was
completed.

\end{document}